\documentclass[a4paper,10pt]{article}
\usepackage[english]{babel}
\usepackage{jheppub}
\usepackage{subfig}
\usepackage{float}
\usepackage{caption}
\usepackage[T1]{fontenc}
\usepackage{graphicx}
\usepackage{epsfig}
\usepackage{pdfpages}
\usepackage{amsmath}
\usepackage{amssymb}
\usepackage{float}
\usepackage{placeins}
\usepackage{braket}
\usepackage{slashed}
\usepackage{mathdots}
\usepackage{lipsum}
\usepackage{bigints}
\usepackage{cleveref}
\usepackage{xcolor}

\usepackage[nodisplayskipstretch]{setspace}

\usepackage{array}

\allowdisplaybreaks[4]

\title{Two-loop form factors for diphoton production in quark annihilation channel with heavy quark mass dependence}

\author[a]{Matteo Becchetti,}
\author[b]{Roberto Bonciani,}
\author[c]{Leandro Cieri,}
\author[c]{Federico Coro,}
\author[b]{Federico Ripani}
\affiliation[a]{Dipartimento di Fisica, Universit\`a di Torino and INFN Sezione di Torino, Via Pietro Giuria 1, I-10125 Torino, Italy}
\affiliation[b]{Dipartimento di Fisica, Universit\`a di Roma “La Sapienza” and
INFN Sezione di Roma, Piazzale Aldo Moro 2, I-00185 Roma, Italy}
\affiliation[c]{Instituto de Fisica Corpuscular, Universitat de Valencia – Consejo Superior de Investigaciones Cientificas, Parc Cientific, E-46980 Paterna, Valencia, Spain}
\emailAdd{matteo.becchetti@unito.it}
\emailAdd{roberto.bonciani@roma1.infn.it}
\emailAdd{fcoro@ific.uv.es}
\emailAdd{federico.ripani@uniroma1.it}

\abstract{We present the computation of the two-loop form factors for diphoton production in the quark annihilation channel. These quantities are relevant for the NNLO QCD corrections to diphoton production at LHC recently presented in \cite{Becchetti:ta}. The computation is performed retaining full dependence on the mass of the heavy quark in the loops. The master integrals are evaluated by means of differential equations which are solved exploiting the generalised power series technique.}

\newcommand{\beq}{\begin{equation}}
\newcommand{\be}{\begin{equation}
\newcommand{\ee}{\end{equation}}}
\newcommand{\eeq}{\end{equation}}
\newcommand{\nn}{\nonumber}
\newcommand{\bea}{\begin{eqnarray}}
\newcommand{\eea}{\end{eqnarray}}
\newcommand{\bfig}{\begin{figure}}
\newcommand{\efig}{\end{figure}}
\newcommand{\bc}{\begin{center}}
\newcommand{\ec}{\end{center}}

\allowdisplaybreaks[4]
\date{}

\begin{document}
\begin{flushright}
IFIC/23-32\\
FTUV-23-0808.6381
\end{flushright}
\maketitle
\flushbottom

\section{Introduction}

The production of photon pairs (diphotons) at the Large Hadron Collider (LHC) is a very relevant process for phenomenological studies in the context of the Standard Model (SM) \cite{ATLAS:2021mbt,ATLAS:2017cvh,CMS:2014mvm,ATLAS:2012fgo} and in the search for new physics \cite{ATLAS:2023hbp,ATLAS:2023meo,ATLAS:2022abz,CMS:2019pov,CMS:2016kgr,ATLAS:2017ayi}.
In particular, diphoton final states are highly relevant for Higgs boson studies \cite{ATLAS:2022fnp,CMS:2022wpo,CMS:2021kom,CMS:2020xrn,ATLAS:2018hxb,CMS:2014afl,ATLAS:2014cnc,ATLAS:2014yga} (and played a crucial role in its discovery \cite{ATLAS:2012yve,CMS:2012qbp}), as they constitute an irreducible background for a Higgs boson decaying into two photons.

Due to its physical relevance, the study of diphoton production requires dedicated and accurate theoretical calculations, in particular including QCD radiative corrections at high perturbative orders. The state of the art for diphoton production is represented by the \textit{next-to-next-to-leading order} (NNLO) QCD corrections \cite{Catani:2011qz,Campbell:2016yrh,Catani:2018krb,Schuermann:2022qdm} to the Born subprocess (which proceeds \textit{via} quark annihilation $q\bar{q} \rightarrow \gamma\gamma$). The relevant scattering amplitudes in the completely massless case have been known in the literature for some time \cite{Ametller:1985di,Parke:1986gb,Dicus:1987fk,Barger:1989yd,Mangano:1990by,Bern:1994fz,Signer:1995np,Balazs:1997hv,DelDuca:1999pa,Anastasiou:2002zn,DelDuca:2003uz}.

More recently, scattering amplitudes belonging to higher orders in the strong coupling $\alpha_{s}$ (i.e. beyond the NNLO) have become available (in the massless case): the three-loop matrix element \cite{Caola:2022dfa}; the two-loop scattering amplitudes for a photon pair in association with one jet in the leading colour approximation \cite{Chawdhry:2020for,Agarwal:2021grm,Chawdhry:2021mkw}, and, very recently, the full colour case \cite{Agarwal:2021vdh}. The two-loop scattering amplitudes for diphoton production in gluon fusion \cite{Bern:2001df} together with the recent computation of diphoton production in association with one jet \cite{Chawdhry:2021hkp} at NNLO emphasise that all the building blocks are in place for the \textit{next-to-next-to-next-to-leading order} (N$^3$LO) \textit{massless} calculation. However, the implementation of slicing subtraction methods to reach the N$^3$LO accuracy could be challenging in this case, due to the presence of a high number of particles in the final state and the use of a photon isolation prescription. More clearly, the NNLO calculation of the diphoton cross section in association with one jet at small diphoton transverse momentum could be very CPU demanding.

In the massive case, the first non-trivial corrections appear at the NNLO, through the inclusion of top quark loops and top quark radiation\footnote{The inclusion of the massive b-quark contribution it is also possible in this context but it is often not considered in the literature.}. The mass effects of the so-called \textit{box} contribution ($gg \rightarrow \gamma\gamma$) were discussed in ref. \cite{Campbell:2016yrh} (together with partial N$^3$LO contributions). Due to the large gluon luminosity at the LHC, the size of the box contribution is of the order of the Born subprocess $q\bar{q} \rightarrow \gamma\gamma$. It is therefore of interest to calculate the corrections of the following perturbative order (i.e. N$^3$LO contributions). Regarding the gluon fusion channel only, the simplest approach (which captures very sizable components) is to consider the NLO QCD corrections to the box contribution, since these form a gauge invariant subset \cite{Bern:2001df} of the whole N$^3$LO gluon fusion channel. In this context, two recent papers have shown the impact of these massive NLO QCD corrections on the gluon fusion channel \cite{Maltoni:2018zvp,Chen:2019fla}.

Considering the full NNLO accuracy, there are still three missing ingredients that were not available or not presented together in previous phenomenological studies in the literature: i) the massive one-loop real-virtual contribution ($q\bar{q} \rightarrow \gamma\gamma g$ and $q(\bar{q}) g\rightarrow \gamma\gamma q(\bar{q})$), the double real radiation of top quarks ($q\bar{q} \rightarrow \gamma\gamma t \bar{t}$ and $gg \rightarrow \gamma\gamma t \bar{t}$) and the two-loop virtual corrections to the Born sub-process $q\bar{q} \rightarrow \gamma\gamma$.

In this paper we report on the computation of the two-loop form factors for diphoton production in the quark annihilation channel, where the full dependence on the heavy quark mass, which appear in the loops, is retained. This contribution is included in the recently presented phenomenological study of full massive NNLO QCD corrections to diphoton production \cite{Becchetti:ta}.

The two-loop form factors have been computed employing standard techniques for scattering amplitude calculations. We considered
the partonic process $q\bar{q} \to \gamma\gamma$ and the relative 
two-loop Feynman diagrams, which contain a massive heavy quark loop, as shown in figure~\ref{fig:typesdiagrams}. The associated amplitude is decomposed into a combination of tensors multiplied by scalar form factors, as described in \cite{Peraro:2020sfm}, and the scalar integrals appearing in the expressions of the form factors are written in terms of a basis of master integrals (MIs). The decomposition in terms of MIs is performed using Identity-by-Parts (IBPs) relations \cite{Chetyrkin:1979bj,Chetyrkin:1981qh}, via the Laporta Algorithm \cite{Laporta:2001dd}, implemented in the computer code\footnote{Other available implementations are described in \cite{Anastasiou:2004vj,Studerus:2009ye,vonManteuffel:2012np,Lee:2012cn,Lee:2013mka,Smirnov:2008iw,Smirnov:2014hma}} \texttt{KIRA} \cite{Maierhofer:2017gsa,Klappert:2020nbg}.

The MIs relevant for this process have been computed by means of the differential equations method \cite{Kotikov:1990kg,Kotikov:1991pm,Bern:1993kr,Remiddi:1997ny,Gehrmann:1999as,Argeri:2007up,Henn:2013pwa,Henn:2014qga}. While the integrals associated to the planar topologies have already been studied in the literature \cite{Becchetti:2017abb,Caron-Huot:2014lda,Bonciani:2003te,Aglietti:2004tq,Bonciani:2003hc,Bonciani:2007eh,Bonciani:2008ep}, a complete computation for the non-planar ones is still not available. Indeed, while the planar MIs admit an analytic solution in terms of Multiple Polylogarithmic functions (MPLs) \cite{Goncharov:2001iea,Goncharov:1998kja,Goncharov2001,Remiddi:1999ew,Vollinga:2004sn,Duhr:2019tlz}, it is known that the functional space for the analytic solution of the non-planar double-box family \cite{Adams:2017tga} contains elliptic integrals \cite{brown2013multiple,Broedel:2014vla,Broedel:2017kkb,ManinModular,Brown:mmv,Adams:2014vja,Frellesvig:2021hkr,Frellesvig:2023iwr,Gorges:2023zgv,Duhr:2022pch,Duhr:2022dxb,Bourjaily:2022bwx}. In recent years, a big effort has been devoted to the understanding of the analytic structure of Feynman integrals which do not admit an expression in terms of MPLs. However,
even in the cases in which an analytic solution in closed form is available, the numerical evaluation of the functions associated to such solution can be extremely challenging for phenomenological applications \cite{Abreu:2022vei,Abreu:2022cco}.

In order to be able to overcome the issues previously described, we choose to exploit the generalised power series method \cite{Pozzorini:2005ff,Aglietti:2007as,Bonciani:2018uvv,Lee:2017qql,Mandal:2018cdj,Moriello:2019yhu,Bonciani:2019jyb} to solve the systems of differential equations associated to the MIs. This technique, currently implemented in two public computer codes \cite{Hidding:2020ytt,Armadillo:2022ugh}, has recently attracted a lot of interest due to its wide range of applicability and it has been successfully employed in several phenomenological applications \cite{Becchetti:2020wof,Bonciani:2021zzf,Becchetti:ta,Bonciani:2022jmb}.
Specifically, in the calculation reported in this paper, we used the software \texttt{DiffExp} \cite{Hidding:2020ytt} to obtain a semi-analytic solution for the MIs. 
 
The paper is organised as follows. In section \ref{sec:setup}, we describe the general setup for the computation and we set the context in which the two-loop form factors, presented in this paper, are relevant. We discuss the form factors computation along with their UV singularity structure and the renormalisation procedure. In section \ref{sec:mis} instead we report on the MIs calculation. We describe the different integral families that appear in the computation and the approach that we used to solve the differential equations, along with a brief analysis on the geometry underlying the actual analytic solution. 

Moreover, as ancillary material attached to this paper, we furnish the analytic expressions of the finite reminder for the form factors, alongside with \texttt{Mathematica} files which allows for a standalone evaluation of the MIs with \texttt{DiffExp}.

\section{Computational setup and amplitude structure}
\label{sec:setup}

In this paper, we consider the two-loop form factors for diphoton production in the quark annihilation channel with a heavy quark loop. At the partonic level, the scattering amplitude proceeds as the Born subprocess:
\begin{equation}
q(p_1)+\overline{q}(p_2)\rightarrow \gamma(p_3)+\gamma(p_4).
\end{equation}
The kinematics for this process is described by the Mandelstam variables\footnote{For our computations we use the metric of \cite{Veltman:1994wz}.}
\begin{equation} \label{eq:mandelstam}
s=-(p_1+p_2)^2, \; t=-(p_1-p_3)^2, \; u=-(p_2-p_3)^2, \; \text{with} \; s+t+u = 0,
\end{equation}
\begin{figure}[h!]
\includegraphics[width=\textwidth]{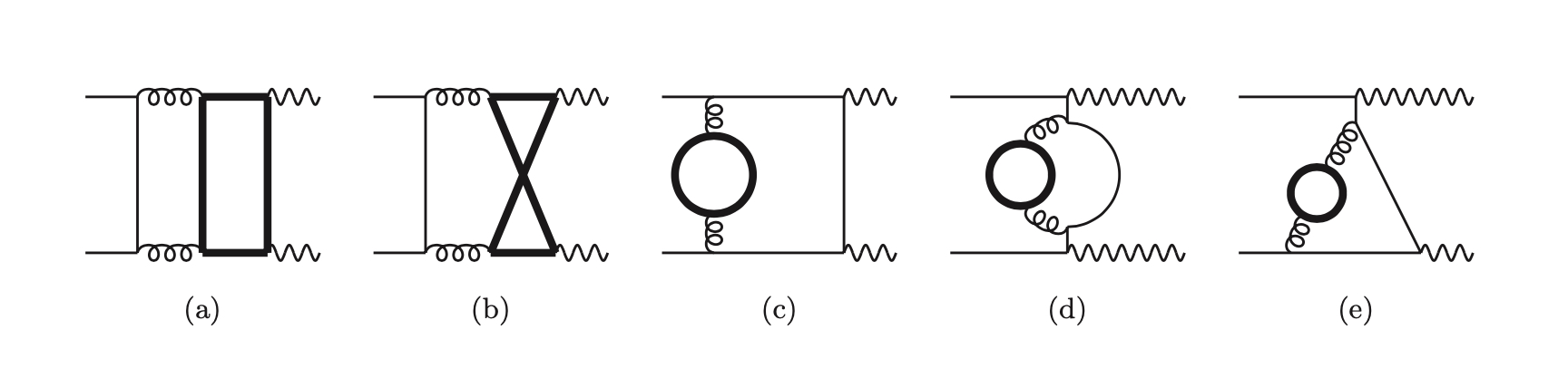}
%  \end{center}
  \caption{Representative set of two-loop diagrams with internal heavy-quark loops, which contribute at NNLO QCD corrections to diphoton production in the quark annihilation channel. Thin black lines represents light quarks, thick black lines heavy quarks, curly lines gluons and curby lines photons.}
  \label{fig:typesdiagrams}
\end{figure}
where the external particles are on-shell, i.e. $p_i^2 = 0$, and we indicate with $m_t^2$ the heavy-quark squared mass\footnote{For the rest of this paper we will refer to the heavy quark as top quark. We note however that our formulas are general and they can be evaluated with a different value of the heavy quark mass.}. In order to obtain the scattering amplitude, we generated the relevant Feynman diagrams using the $\tt{FeynArts}$ package \cite{Hahn:2000kx}. We found a total number of 14 diagrams contributing to the amplitude, the representative ones are shown in fig. \ref{fig:typesdiagrams}. We write the scattering amplitude in terms of form factors, which are decomposed into a basis of 72 MIs exploiting IBPs reduction \cite{Chetyrkin:1979bj,Chetyrkin:1981qh,Laporta:2001dd,Lee:2012cn,Lee:2013mka,Smirnov:2008iw,Smirnov:2014hma,Studerus:2009ye,vonManteuffel:2012np,Peraro:2019svx,Maierhofer:2017gsa}, as implemented in the software $\tt{Kira}$ \cite{Maierhofer:2017gsa}.

The MIs contributing to this process can be described by three different scalar integral topologies (modulo exchange of the two final photons). Specifically, the MIs for the Feynman diagrams (a) and (b) in fig. \ref{fig:typesdiagrams} are associated to the integral families PLA and NPL, respectively, as defined in section \ref{sec:mis}. Similarly, the MIs for the diagrams (c), (d) and (e) can be grouped into one scalar integral family, PLB, also defined in section \ref{sec:mis}. The MIs of the families PLA and PLB were already known in the literature \cite{Becchetti:2017abb}. Regarding the non-planar topology NPL, while most of the MIs have already been studied \cite{Becchetti:2017abb,Bonciani:2007eh,Bonciani:2008ep,Aglietti:2006tp,Anastasiou:2006hc,vonManteuffel:2017hms,Bonciani:2018uvv}, the double-box top-sector have not been considered in the literature yet, and therefore its computation represents an original result by itself.

For this project, we performed an independent calculation of all the MIs by means of the differential equations method \cite{Kotikov:1990kg,Kotikov:1991pm,Bern:1993kr,Remiddi:1997ny,Gehrmann:1999as,Argeri:2007up,Henn:2013pwa,Henn:2014qga}. In particular we solved the system of differential equations semi-analytically exploiting the generalised power series expansion technique, as described in \cite{Moriello:2019yhu} and implemented in the software $\tt{DiffExp}$ \cite{Hidding:2020ytt}.

The two-loop amplitudes computed in this paper constitute a necessary ingredient of the recently presented full massive NNLO QCD corrections to diphoton production at hadron colliders \cite{Becchetti:ta}. We anticipate here, that our two-loop form factors, taking into account the full dependence on the top quark mass, are finite (free of IR divergences after UV renormalization). Therefore, they can be included directly (without the use of any IR regularization prescription) in any numerical implementation of the NNLO cross section. In the following paragraph we will illustrate the precedent statement with a specific example based on the $q_T$-subtraction method \cite{Catani:2007vq,Catani:2013tia} (which can easily be extended to any other subtraction method).

To this end, we consider the following scattering process,
\begin{equation}
    h_1 + h_2 \rightarrow \gamma\gamma
\end{equation}
where $h_1$ and $h_2$ are the colliding hadrons. At the NNLO, the cross-section for this process can be computed using the $q_T$-subtraction method \cite{Bozzi:2005wk,Catani:2007vq,Catani:2013tia} as follows
\begin{equation} \label{eq:NNLOcs}
    d\sigma^{\gamma\gamma}_{\rm NNLO} = \mathcal{H}^{\gamma\gamma}_{\rm NNLO} \otimes d\sigma^{\gamma\gamma}_{\rm LO} + \left[d\sigma^{\gamma\gamma+{\rm jets}}_{\rm NLO} - d\sigma^{CT}_{\rm NLO}\right].
\end{equation}
The terms inside the square brackets $d\sigma^{\gamma\gamma+jets}_{\rm NLO}$ and $d\sigma^{CT}_{\rm NLO}$ represent the cross section for diphoton plus jet production at NLO \cite{DelDuca:2003uz} and the corresponding counterterm, needed to cancel the associated singularities in the small-$q_T$ limit. The coefficient function $\mathcal{H}^{\gamma\gamma}_{\rm NNLO}$ (defined in \cite{Bozzi:2005wk}) is the so-called hard-virtual function and it includes the one-loop and two-loop corrections to the Born subprocess. This object admits a perturbative expansion in terms of the strong coupling $\alpha_S$:
\begin{equation} \label{eq:hf_exp}
    \mathcal{H}^{\gamma\gamma} =1 + \frac{\alpha_S}{\pi}\mathcal{H}^{\gamma\gamma}_{\rm NLO} + \left(\frac{\alpha_S}{\pi}\right)^2 \mathcal{H}^{\gamma\gamma}_{\rm NNLO} + \cdots\,.
\end{equation}

In our particular case (diphoton production) the one-loop contribution to eq. \eqref{eq:hf_exp} was calculated in \cite{Balazs:1997hv}, while the massless two-loop contribution was first calculated in \cite{Anastasiou:2002zn} and later in \cite{Caola:2022dfa}. The explicit expressions of the hard virtual factor (computed with the previous massless one- and two-loop amplitudes) in the hard resummation scheme are given in Appendix A of ref. \cite{Catani:2013tia}. The inclusion of the new two-loop massive form factors proceeds by simple addition to that of the massless case \cite{Anastasiou:2002zn}. After regularising the IR divergences (e.g. in the \textit{in the hard resummation scheme}) \cite{Catani:2013tia} present in the massless two-loop amplitude \cite{Anastasiou:2002zn}, our massive two-loop contribution can be added directly to the massless finite remainder.

\label{sec:amp}

After generating all the relevant Feynman diagrams for the process, the amplitude $\mathcal{A}_{q\bar{q},\gamma\gamma}$ has been decomposed in terms of form factors \cite{Peraro:2020sfm}, and the UV singularities have been regularised in dimensional regularisation. The expression obtained has been used to compute the NNLO corrections to the hard function $\mathcal{H}^{\gamma\gamma}$ coming from two-loop diagrams which involve a massive top quark loop. Specifically, from the knowledge of the finite remainder $\mathcal{A}^{(\text{fin})}_{q\bar{q},\gamma\gamma}$ of the amplitude, we can obtain the hard function from the all-orders relation \cite{Catani:2013tia}:
\begin{equation}
    \mathcal{H}^{\gamma\gamma} = \frac{\vert \mathcal{A}^{(\text{fin})}_{q\bar{q},\gamma\gamma} \vert^2}{\vert \mathcal{A}_{q\bar{q},\gamma\gamma}^{(0)} \vert^2},
\end{equation}
where $\mathcal{A}_{q\bar{q},\gamma\gamma}^{(0)}$ is the Born-level amplitude for this process:
\begin{equation}
    \vert \mathcal{A}_{q\bar{q},\gamma\gamma}^{(0)} \vert^2 = 128 \, \pi^{2} \, \alpha_{em}^2 \, Q_q^4 \, N_c\left(\frac{t}{u} + \frac{u}{t}\right),
\end{equation}
with $\alpha_{em}$, the QED coupling, $Q_q$ is the electric charge of the incoming quarks and $N_c$ is the number of colours. We also performed a sum over initial and final polarisations and initial colours.

This massive hard-virtual coefficient represents the last missing ingredient necessary to perform a NNLO phenomenological study, for diphoton production at the LHC, which takes into account the complete dependence on the top quark mass \cite{Becchetti:ta}.

\subsection{Form factors}

The bare scattering amplitude can be written\footnote{For the sake of simplicity we are omitting color indices on the left side of Eq. \eqref{eq:amp}} as
\begin{equation}
\label{eq:amp}
\mathcal{A}_{q\bar{q},\gamma\gamma} = \alpha_{em}\,\delta_{ij}\,\epsilon^{\ast\mu}_{\lambda_3}(p_3)\epsilon^{\ast\nu}_{\lambda_4}(p_4)\overline{v}_{s_2}(p_2)\mathcal{A}_{\mu\nu}(s,t,u,m_t^2)u_{s_1}(p_1), 
\end{equation}
where $\delta_{ij}$ is the Kronecker delta function with $i$ and $j$ the color indices of the incoming $q\bar{q}$ pair, $\epsilon^{\ast\mu}_{\lambda_3}(p_3)$ and $\epsilon^{\ast\nu}_{\lambda_4}(p_4)$ are external photon polarisation vectors and $\overline{v}_{s_2}(p_2)$, $u_{s_1}(p_1)$ the quark spinors.

The amplitude \eqref{eq:amp} can be further decomposed in terms of a set of four independent tensors\footnote{In general one has to consider a fifth form factor, however, for the corrections we are considering in this paper it can be chosen in such a way that the tensor is proportional to $D-4$ and the form factor is finite after UV renormalisation. Hence, it vanishes in $D=4$ and it can be neglected for phenomenological considerations.} \cite{Peraro:2020sfm} which are built using external momenta and polarisation vectors:
\begin{equation} \label{eq:f_f}
\mathcal{A}_{q\bar{q},\gamma\gamma}=\sum_{k=1}^4\mathcal{F}_k \, T_k,
\end{equation}
where the $T_k$ are chosen as:
\begin{equation}
\begin{split}
T_1&=\overline{v}_{s_2}(p_2)\slashed{\epsilon}^{\ast}_{\lambda_3}(p_3)u_{s_1}(p_1)\epsilon^{\ast}_{\lambda_4}(p_4)\cdot p_2, \\
T_2&=\overline{v}_{s_2}(p_2)\slashed{\epsilon}^{\ast}_{\lambda_4}(p_4)u_{s_1}(p_1)\epsilon^{\ast}_{\lambda_3}(p_3)\cdot p_1, \\
T_3&=\overline{v}_{s_2}(p_2)\slashed{p}_3u_{s_1}(p_1)\epsilon^{\ast}_{\lambda_3}(p_3)\cdot p_1\epsilon^{\ast}_{\lambda_4}(p_4)\cdot p_2, \\
T_4&=\overline{v}_{s_2}(p_2)\slashed{p}_3u_{s_1}(p_1)\epsilon^{\ast}_{\lambda_3}(p_3)\cdot\epsilon^{\ast}_{\lambda_4}(p_4).
\end{split}
\end{equation}
The decomposition \eqref{eq:f_f} has been achieved also by enforcing the physical conditions $\epsilon_{\lambda_i}\cdot p_i=0$ for the polarisation vectors, and by choosing as reference vectors for the external photons the condition $\epsilon_{\lambda_3} \cdot p_{2} = \epsilon_{\lambda_4} \cdot p_1 = 0$, which implies for the photons the polarization sums:
\begin{equation}
    \sum_{\lambda_3} \epsilon_{\lambda_3}^{\mu} \epsilon_{\lambda_3}^{*\nu} = g^{\mu \nu} - \frac{p_2^{\mu}p_3^{\nu} + p_3^{\mu}p_2^{\nu}}{p_2 \cdot p_3}, \hspace{1cm}
    \sum_{\lambda_4} \epsilon_{\lambda_4}^{\mu} \epsilon_{\lambda_4}^{*\nu} = g^{\mu \nu} - \frac{p_1^{\mu}p_4^{\nu} + p_4^{\mu}p_1^{\nu}}{p_1 \cdot p_4},
\end{equation}
where $g^{\mu\nu} = \operatorname{diag}(1,1,1,1)$. The coefficients of $T_k$ are the so-called scalar form factors, $\mathcal{F}_k$. These objects are functions of the kinematic invariants of the process, and of the space-time dimension $D$, and they can be written in terms of scalar Feynman integrals. Their expression can be obtained by applying a set of projectors, $\left\{\mathcal{P}_k\right\}$, $k=1,2,3,4$, to the amplitude $\mathcal{A}_{q\bar{q},\gamma\gamma}$:
\begin{equation}
\mathcal{F}_k= \operatorname{tr} \left\{ \sum_{s_1,s_2,\lambda_3,\lambda_4}\mathcal{P}_k \, \mathcal{A}_{q\bar{q},\gamma\gamma} \right\},
\end{equation}
where
\begin{align}
\mathcal{P}_1&=\frac{1}{(D-3)t}\left[\frac{u}{2s^2}T_1^{\dag}-\frac{u}{2 s^2 t}T_3^{\dag}\right], \nn \\
\mathcal{P}_2&=\frac{1}{(D-3)t}\left[\frac{u}{2 s^2 t}T_3^{\dag}+\frac{u}{2 s^2}T_2^{\dag}\right], \nn \\
\mathcal{P}_3&=\frac{1}{(D-3)t}\left[\frac{\left(D u^2 - 4 s t\right)}{2 s^2 u t^2}T_3^{\dag}-\frac{u}{2 s^2 t}T_1^{\dag}+\frac{u}{2 s^2 t}T_2^{\dag}+\frac{(t-s)}{2 s u t}T_4^{\dag}\right], \nn \\
\mathcal{P}_4&=\frac{1}{(D-3)t}\left[\frac{(t-s)}{2 s u t}T_3^{\dag}+\frac{1}{2 u}T_4^{\dag}\right].
\end{align}
The form factors $\mathcal{F}_k$ admit the following perturbative expansion:
\begin{equation}
\mathcal{F}_k=\mathcal{F}_k^{(0)}+\left(\frac{\alpha_S}{\pi}\right)\mathcal{F}_k^{(1)}+\left(\frac{\alpha_S}{\pi}\right)^2\mathcal{F}_k^{(2)}+\cdots \, ,
\end{equation}
where $\alpha_S$ is the strong coupling constant. At leading order we have:
\begin{align}
\mathcal{F}_1^{(0)} &= - 4 \pi \alpha_{em}\delta_{ij} Q_q^2 \, \frac{1}{t} \, , \nn \\
\mathcal{F}_2^{(0)} &=  4 \pi \alpha_{em}\delta_{ij} Q_q^2 \, \frac{1}{t} \, , \nn \\
\mathcal{F}_3^{(0)} &=  4 \pi \alpha_{em}\delta_{ij} Q_q^2 \, \frac{2}{tu} \, , \nn \\
\mathcal{F}_4^{(0)} &= - 4 \pi \alpha_{em}\delta_{ij} Q_q^2 \, \frac{(t-u)}{tu} \, .
\end{align}
The massive corrections, we are interested in, appear starting from the two-loop order. Therefore, they affect only the term $\mathcal{F}_k^{(2)}$. For the rest of this paper we will focus just on this contribution, which we will refer to as $\mathcal{F}_{k,\texttt{top}}^{(2)}$. We generated all the relevant Feynman diagrams for this contribution using \texttt{FeynArts} \cite{Hahn:2000kx}, and we found 14 different two-loop Feynman diagrams, which can be grouped into five different categories, as depicted in fig. \ref{fig:typesdiagrams}. We used \texttt{FORM} \cite{Kuipers:2012rf,Ruijl:2017dtg} to apply the projectors to the Feynman diagrams and perform the Dirac algebra. The form factors, then, were expressed as a linear combinations of 72 MIs, $\left\{\mathcal{J}_i\right\}$, which are defined in Appendix \ref{appendix}.

We find the following expression for the form factors $\mathcal{F}_{k,\texttt{top}}^{(2)}$:
\begin{equation} \label{eq:F2}
\mathcal{F}_{k,\texttt{top}}^{(2)} = 4 \pi \alpha_{em} \delta_{ij} \, C_F \left[Q_{q}^{2}\mathcal{F}_{k,\texttt{top};0}^{(2)} + Q_t^2 \mathcal{F}_{k,\texttt{top};2}^{(2)}\right] \, ,
\end{equation}
where $C_F=\frac{N_c^2-1}{2N_c}$ is the Casimir of the fundamental representation of $SU(N_c)$ and $Q_{t}$ is the electric charge of the top quark running in the loop. The two contributions $\mathcal{F}_{k,\texttt{top};0}^{(2)}$ and $\mathcal{F}_{k,\texttt{top};2}^{(2)}$ are indeed related to different powers of the top electric charge $Q_t$. The first contribution $\mathcal{F}_{k,\texttt{top};0}^{(2)}$ is associated to the diagrams (c), (d) and (e) in fig. \ref{fig:typesdiagrams} in which the top quark does not couple to the external photons, while the second contribution $\mathcal{F}_{k,\texttt{top};2}^{(2)}$ comes from the diagrams (a) and (b) where the top quark actually couples with the photon.

We perform our computations in the context of dimensional regularisation. As a consequence, potential ultraviolet (UV) and infrared (IR) singularities can appear in the form factors as poles in the dimensional regulator $\epsilon=(4-D)/2$. However, since the diagrams with a top loop start contributing to the $q\bar{q}$ channel at the two-loop order,  $\mathcal{F}_{k,\texttt{top}}^{(2)}$ does not have IR singularities 
and therefore all the $\epsilon$ poles are of UV origin. Furthermore, $\mathcal{F}_{k,\texttt{top};2}^{(2)}$ is also free of UV divergences and then the UV poles come only from the contribution $\mathcal{F}_{k,\texttt{top};0}^{(2)}$.

\subsection{UV Renormalisation}

We renormalise the bare form factors in a mixed scheme. The external quark fields are renormalised on shell; the strong coupling constant is renormalised in a scheme in which the light-quark contribution is treated in $\overline{\mbox{MS}}$, while the heavy-quark contribution is renormalised at zero momentum. No renormalisation is needed for the top quark mass $m_t^2$ at this order in perturbation theory and the same occurs for the external photon field.

We have, then
\be
\mathcal{F}_{k}^{R} = Z_q \, \mathcal{F}_{k}\left( \alpha_S^B \to Z_{\alpha_S}\alpha_S \right) \, ,
\ee
where the superscript ``R'' stands for ``renormalised''.
The renormalisation factors $Z_q$ and $Z_{\alpha_S}$ admit a perturbative expansion in the strong coupling constant \cite{Barnreuther:2013qvf,Broadhurst:1993mw,Melnikov:2000qh,Mitov:2006xs}:
\begin{align}
Z_{q} &= 1 + \left(\frac{\alpha_S}{\pi}\right)\delta Z_{q}^{(1)}  + \left(\frac{\alpha_S}{\pi}\right) ^2 \delta Z_{q}^{(2)}  + \mathcal{O}(\alpha_S^4) \, , \nn \\
Z_{\alpha_S} &= 1 +\left(\frac{\alpha_S}{\pi}\right)\delta Z_{\alpha}^{(1)}  + \mathcal{O}(\alpha_S^2) \, .
\end{align}
In our renormalisation scheme, we have
\begin{align}
\delta Z_{q}^{(1)} & = 0 \, , \nn \\
\delta Z_{q}^{(2)} & = \pi^{-2\epsilon} \Gamma^2(1+\epsilon) \, \left( \frac{\mu^{2}}{m_t^2} \right)^{2\epsilon} \, C_F N_h T_F \, \left(\frac{1}{16 \epsilon} - \frac{5}{96}\right) \, , \nn \\
\delta Z_{\alpha}^{(1)} & = \delta Z_{\alpha,N_l,\overline{\mathrm{MS}}}^{(1)} + \delta Z_{\alpha,N_h,\mathrm{OS}}^{(1)} \, ,
\end{align}
where
\begin{align}
\delta Z_{\alpha,N_l,\overline{\mathrm{MS}}}^{(1)} & = - \pi^{-\epsilon} e^{- \gamma \epsilon} \frac{1}{2\epsilon} \left( \frac{11}{6} C_A - \frac{1}{3} N_l \right) \, , \nn \\
\delta Z_{\alpha,N_h,\mathrm{OS}}^{(1)} & = 
\pi^{-\epsilon} \Gamma(1+\epsilon) \, \left( \frac{\mu^{2}}{m_t^2} \right)^{\epsilon} \, \frac{N_h}{6\epsilon} \, ,
\end{align}
$\mu$ is the renormalisation scale,  $N_l$ is the number of light flavours, in this case $N_l=5$, $N_h$ is the number of heavy quarks, in this case $N_h=1$, the quantity $C_A = N_c$ is the  Casimir of the adjoint representation of $SU(N_c)$, and $T_F = \frac{1}2$.

Therefore, the renormalised form factors $\mathcal{F}_{k,\texttt{top}}^{(2) \, R}$ are determined by the following formula:
\begin{equation} \label{eq:f2_ren}
\mathcal{F}_{k,\texttt{top}}^{(2) \, R} = 
\mathcal{F}_{k,\texttt{top}}^{(2)}
+ \delta Z_{q}^{(2)}  \, \mathcal{F}_{k}^{(0)} + \delta Z_{\alpha,N_h,\mathrm{OS}}^{(1)}  \, \mathcal{F}_{k}^{(1)} \, .
\end{equation}

\section{Master Integrals Computation} \label{sec:mis}

In this section we discuss the details of the MIs computation. Specifically, we define the three scalar integral topologies which describe all the MIs that appear in the amplitude (modulo exchange of final photons). The MIs are computed through the differential equations method. The system of differential equations associated to the MIs is solved semi-analytically employing the generalised power series expansion technique, as described in \cite{Moriello:2019yhu} and implemented in the software $\tt{DiffExp}$ \cite{Hidding:2020ytt}. Finally, we show that some of the MIs that appear in the computation admit an analytic solution in terms of elliptic functions.

The system of differential equations for the planar families, PLA and PLB, is in canonical form, while for the non-planar family, NPL, we have canonical differential equations just for the sectors whose analytic solution could be given in terms of MPLs. We notice that the bases of MIs in which we solve the differential equations, which we will refer in the following as $\vec{f}$ for the planar families and $\vec{g}$ for the non-planar one, and the one in which we write the form factors are different. We choose this approach in order to avoid square roots of the kinematic invariants in the expressions of the form factors. In the ancillary material attached to this paper, we furnish the rotation matrices from the bases $\vec{f}$ and $\vec{g}$ to the basis $\left\{\mathcal{J}_i\right\}$ defined in Appendix \ref{appendix}.

\subsection{General Setup}

The three integral families which describe the relevant MIs for this process are defined as follows:
\begin{equation}
\mathcal{I}_{\texttt{topo}}(n_1,...,n_9) = \int \frac{\mathcal{D}k_1\mathcal{D}k_2}{D_{1}^{n_1}D_{2}^{n_2}D_{3}^{n_3}D_{4}^{n_4}D_{5}^{n_5}D_{6}^{n_6}D_{7}^{n_7}D_{8}^{n_8}D_{9}^{n_9}}, 
\end{equation}
where \texttt{topo} $\in\left\{\text{PLA}, \text{PLB}, \text{NPL}\right\}$ labels the families and the scalar products $D_1,...,D_9$ are given in table~\ref{fig:topos}. The indices $n_1,\cdots,n_7$ are non-negative while the indices $n_8$ and $n_9$ are non-positive.

The computation is done in dimensional regularization with $D=4-2\epsilon$ dimensions, and our convention for the integration measure is 

\begin{equation}
\mathcal{D}k=\frac{\operatorname{d}^D k}{i \pi^{2-\epsilon}\Gamma(1+\epsilon)} \left( \frac{\mu^2}{m_t^2}  \right)^{-\epsilon} \, ,
\end{equation}

where $\mu$ is the dimensional regularization scale.

\begin{figure}[h!]
\includegraphics[width=\textwidth]{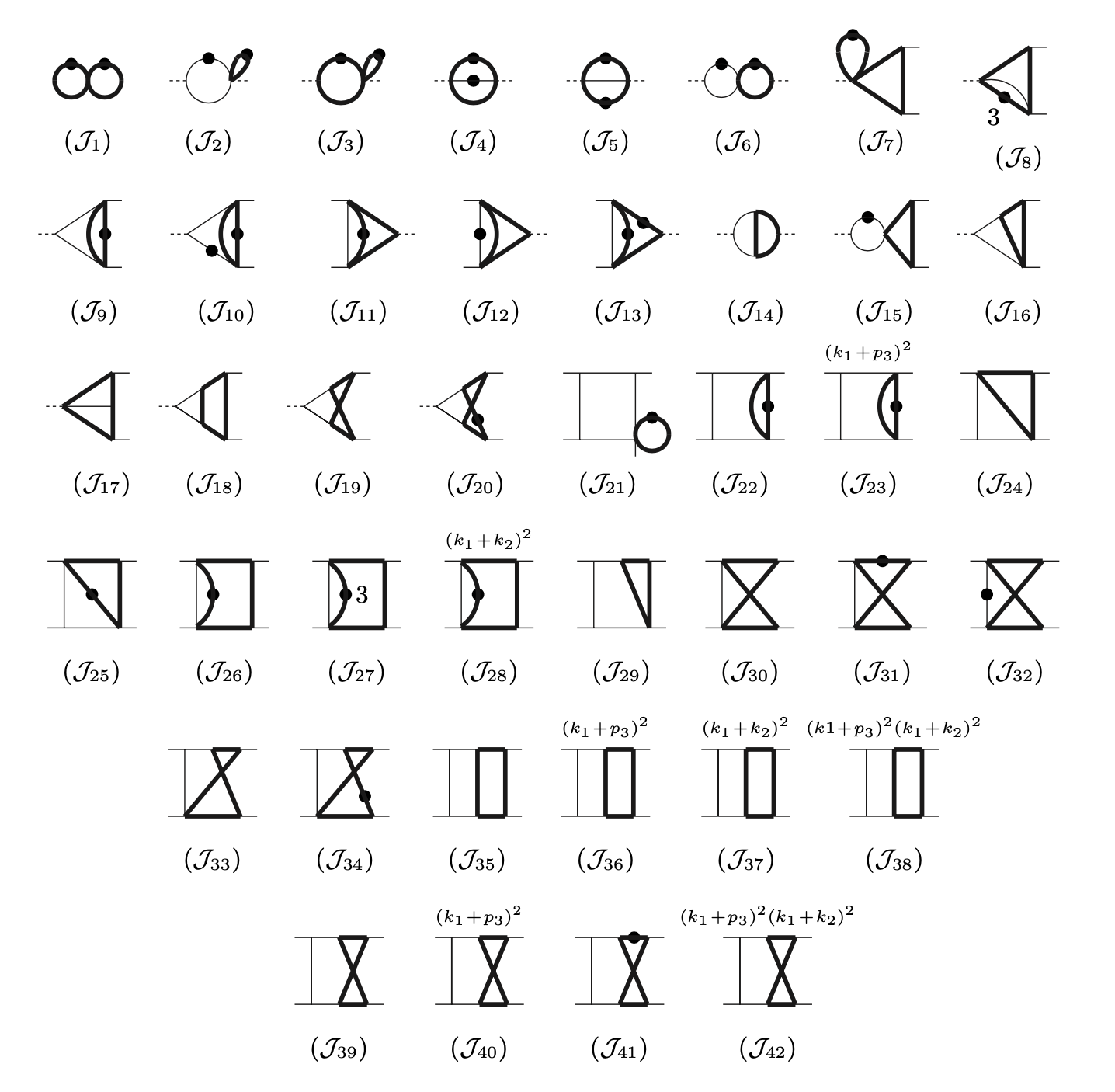}
  \caption{Basis of Master Integrals for the topologies PLA, PLB and NPL. In the figure, the MIs defined with a permutation of the external momenta are not displayed.
  \label{fig:mis}}
\end{figure}

With the definitions given in table~\ref{fig:topos}, all the scalar Feynman integrals appearing in the amplitude can be mapped to one of the three integral families, or families obtained by permutation of the external momenta. The MIs appearing in the form factors are listed in appendix \ref{appendix} and shown in fig.~\ref{fig:mis}.

\begin{table}
\begin{center}
\begin{tabular}{| m{2.5cm} || m{3.5cm} | m{3.5cm} | m{3.5cm} |} 
 \hline
 Denominator & Integral family PLA & Integral family PLB & Integral family NPL \\[5pt]
 \hline
$D_1$ & $k_1^2$       & $k_1^2$       & $k_1^2$\\
$D_2$ & $(k_1-p_1)^2$ & $(k_1-p_1)^2$ & $(k_1-p_1)^2$\\
$D_3$ & $(k_1+p_2)^2$ & $(k_1+p_2)^2$ & $(k_1+p_2)^2$\\
$D_4$ & $k_2^2+m_t^2$ & $k_2^2+m_t^2$ & $k_2^2+m_t^2$\\
$D_5$ & $(k_1+k_2-p_1)^2+m_t^2$ & $(k_2-k_1)^2+m_t^2$ & $(k_1+k_2-p_1)^2+m_t^2$\\
$D_6$ & $(k_1+k_2+p_2)^2+m_t^2$ & $(k_1-p_1+p_3)^2$   & $(k_2-p_1-p_2+p_3)^2+m_t^2$\\
$D_7$ & $(k_1+k_2-p_1+p_3)^2+m_t^2$ & $(k_2+p_1)^2$   & $(k_1+k_2-p_1+p_3)^2+m_t^2$\\
$D_8$ & $(k_1+p_3)^2$ & $(k_2+p_2)^2$ & $(k_1+p_3)^2$\\
$D_9$ & $(k_1+k_2)^2$ & $(k_2+p_3)^2$ & $(k_1+k_2)^2$\\
 \hline
\end{tabular}  \caption{\label{fig:topos}Routing definition for the three scalar integrals families PLA, PLB and NPL.}
\end{center}
\end{table}

\subsection{Planar Families PLA and PLB}
\label{PLA}

The scalar integrals family PLA is described by a set of 32 master integrals $\vec{f}(\vec{x},\epsilon)$\footnote{The definition of the MIs basis exploited to solve the differential equations for the planar families PLA and PLB is explicitly given in the ancillary material attached to the paper.}, where 
\begin{equation}
\vec{x} = \left\{y,z\right\}, \;\;\; y = \frac{s}{m_t^2}, \; z = \frac{t}{m_t^2}
\end{equation}
is the vector of the kinematic invariants with respect to which we derived the differential equations. The system of differential equations for these integrals has been derived in canonical logarithmic form \cite{Henn:2013pwa} in ref. \cite{Becchetti:2017abb}:
\begin{equation} \label{eq:deq_can}
    d\, \vec{f}(\vec{x},\epsilon) = \epsilon\,d\,A(\vec{x}) \vec{f}(\vec{x},\epsilon), 
\end{equation}
where $d$ is the total differential with respect to the kinematic invariants, and the matrix $A(\vec{x})$ is written as linear combinations of logarithms:
\begin{equation}
    A(\vec{x}) = \sum c_i \log (w_i (\vec{x})).
\end{equation}
The $c_i$ represent matrices of rational numbers, while $\mathbf{W}_{PLA} = \left\{w_i(\vec{x})\right\}$
is the alphabet of the solution and it is made by algebraic functions of the kinematic invariants. Specifically, the alphabet is made by the following set of 21 letters:
\begin{align} \label{eq:alph}
    \mathbf{W}_{PLA} = \biggl\{&r_1,r_2,r_3,r_4,\frac{r_4-r_5}{r_4+r_5},y,\frac{r_1-y}{r_1+y},r_4-r_2 y,\frac{r_3+y}{r_3-y},\frac{-r_1 z+r_1+r_5}{- r_1 z+r_1-r_5},\frac{r_5-y z+y}{y-y z},\frac{r_1 z-r_4}{r_1 z+r_4}, \nn \\
&\frac{r_1 z-r_5}{r_1 z+r_5},1-z,z,y+z,\frac{r_2-z}{r_2+z},-\frac{r_5-y z+y+2 z}{r_5-y z+y-2 z^2},\frac{r_4 z+r_4-r_5 z}{r_4 z+r_4+r_5 z},-\frac{-r_2+r_3+y+z}{r_2+r_3+y+z},\nn \\
&\frac{-r_2 y (z-4)+r_2 z (r_2-z+4)+r_4 (z-4)}{r_2 y (z-4)+r_2 z (r_2+z-4)-r_4 (z-4)} \biggr\}, 
\end{align}
where $r_1,\cdots,r_5$ are a set of square roots of the kinematic invariants:
\begin{equation}
\begin{split} \label{eq:roots_PLA}
r_1&= \sqrt{(y-4) y}, \hspace{0.5cm} r_2= \sqrt{(z-4) z}, \hspace{0.5cm} r_3= \sqrt{y (y+4)}, \hspace{0.5cm} r_4= \sqrt{y z (y (z-4)-4 z)},\\
r_5&= \sqrt{y \left(y (z-1)^2-4 z^2\right)}.
\end{split}
\end{equation}
The knowledge of the logarithmic canonical form of the differential equations \eqref{eq:deq_can}, together with the alphabet structure \eqref{eq:alph} would allow us, in principle, to obtain a fully analytic representation for the system of the MIs. However, the presence of the set of square roots given in Eq. \eqref{eq:roots_PLA} makes the achievement of such analytic expression non trivial. Indeed, these square roots are not simultaneously rationalizable. As a consequence, in order to obtain a fully analytic representation of the solution one would have to exploit symbol level techniques \cite{Goncharov:2010jf,Duhr:2011zq}. For the purpose of this project, we found that the semi-analytic evaluation, which we achieved exploiting a generalised power series expansion method, was sufficient to perform phenomenological studies.

The boundary conditions for the system are provided in the origin of kinematic variables $y=z=0$, where all the MIs vanish except for the two masters $f_1$ and $f_2$, for which we use the following analytical expressions:
\begin{align}
        f_1&=\epsilon^2\,\mathcal{J}_1=1, \nn \\
        f_2&(y)=-y\, \epsilon^2\, \mathcal{J}_2(y)=-\left(-y\right)^{-\epsilon }\frac{ \Gamma (1-\epsilon )^2}{\Gamma (1-2 \epsilon )},
\label{f1f2}
\end{align}
where $\mathcal{J}_1$ and $\mathcal{J}_2$ are the pre-canonical masters shown in fig.~\ref{fig:mis} and defined in Appendix \ref{appendix}.

Regarding the second planar family, PLB, we observe that we do not need to set up a system of differential equations for it. Indeed, every master integral of this family, except $\mathcal{J}_{21}$, is equal to one of the MIs of the two other families (modulus permutations). For $\mathcal{J}_{21}$, by integrating analytically its differential equation, we obtain the following analytical expression:

\begin{align}
\mathcal{J}_{21}(y,z) =& \frac{1}{y\,z}\left[\frac{4}{\epsilon^3} \, - \, \frac{2 (\log (-y)+\log (-z))}{\epsilon^2} \, - \, \frac{5 \pi^2-6 \log (-y) \log (-z)}{3 \epsilon} \, + \, \right. \nn \\
& + 2 \text{Li}_3\left(-\frac{z}{y}\right)-2 \log (-z) \text{Li}_2\left(-\frac{z}{y}\right)+2 \log (-y) \text{Li}_2\left(-\frac{z}{y}\right) \nn \\
&- \log (-y) \log ^2(-z)+ \log (y) \log ^2(-z)- \log ^2(-z) \log (y+z) \\
&- \log ^2(-y) \log (y+z) -2 \log (-y) \log (y) \log (-z) \nn \\
&+2 \log (-y) \log (-z) \log (y+z) - \pi ^2 \log (y+z) \nn \\ 
&+ \log ^2(-y) \log (y) +\frac{\pi ^2}{3} \log (-y)+ \pi ^2 \log (y) \nn \\
&\left. \quad+\frac{1}{3}\log ^3(-z)+\frac{4}{3} \pi ^2 \log (-z)-10 \zeta (3) \right] \; + \; \mathcal{O}(\epsilon). \nn
\end{align}
This formula was cross checked analytically with \cite{Henn_2015} and numerically with \texttt{AMFlow} \cite{Liu_2023}.

\subsection{Non-Planar Family NPL}

\begin{figure}[t] 
\begin{center}
\includegraphics[width=7 cm]{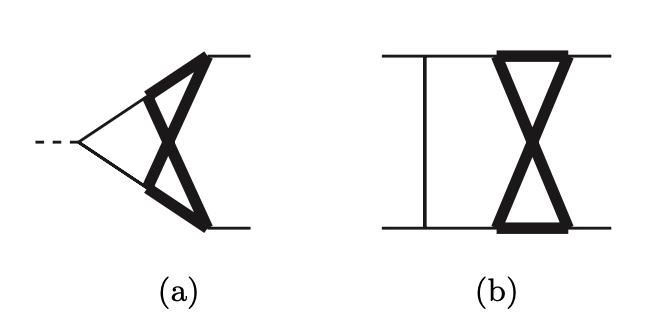}
\end{center}
\caption{Diagrams representing the two elliptic sectors in the non planar topology NPL. Thin lines represent massless particles, while thick lines massive particles.} \label{fig:elliptic_sec}
\end{figure}

The non-planar scalar integrals family, NPL, is more complicated to study with respect to the two planar families. The total number of MIs $\vec{g}$ \footnote{The definition of the MIs basis exploited to solve the differential equations for the non-planar family NPL is explicitly given in the ancillary material attached to the paper.} associated to this topology is 36. The system of differential equations can be divided in two different subsets:
\begin{itemize}
    \item \textbf{(I) Canonical logarithmic}: the subset of MIs whose differential equations is in canonical logarithmic form;
    \item \textbf{(II) Elliptic sectors}: this subset contains MIs whose analytic solution involve elliptic functions and it is not written in $\epsilon$-factorized form.
\end{itemize}
The system of differential equations for the subset (I) has been put in canonical logarithmic form. The alphabet $\mathbf{W}_{NPL}$ for this subset is described by the following 30 letters:
\begin{align}
  \mathbf{W}_{NPL} = \biggl\{&q_1,q_3,q_4,q_5,q_6,-\frac{q_4-2 q_8}{q_4+2 q_8},\frac{q_5-2 q_8}{q_5+2 q_8},-\frac{q_9-2}{q_9+2},q_9,-\frac{q_7 y+q_5}{q_5-q_7 y},y,-\frac{q_1+y}{y-q_1}, \nn \\
  &q_4-q_2 y,-\frac{q_3+y}{y-q_3},\frac{q_1 z-q_4}{q_1 z+q_4},-\frac{q_6-q_7 z}{q_7 z+q_6},z,y+z,y+z+4,-\frac{q_2+z}{z-q_2},\frac{q_1 y+q_1 z+q_5}{q_1 y+q_1 z-q_5}, \nn \\
  &\frac{q_2 y+q_2 z+q_6}{q_2 y+q_2 z-q_6},\frac{-q_5+y^2+y z}{q_5+y^2+y z},\frac{q_3 y-q_7 y+q_3 z}{q_3 y+q_7 y+q_3 z},\frac{-q_6+y z+z^2}{q_6+y z+z^2},\frac{q_6+y z+z^2}{-q_6+y z+z^2}, \nn \\
  &-\frac{q_7+y+z}{-q_7+y+z},-\frac{q_2+q_3+y+z}{-q_2+q_3+y+z},-\frac{q_8 y z-4 q_8 y+q_6 q_9 y+q_8 z^2-4 q_8 z-2 q_8 q_9 z}{q_8 y z-4 q_8 y+q_6 q_9 y+q_8 z^2-4 q_8 z+2 q_8 q_9 z}, \nn \\
  &\frac{-q_2 y-q_2 z+q_4+z^2}{q_2 y+q_2 z-q_4+z^2}\biggr\}
\end{align}
where $q_1,\cdots,q_9$ are a set of square roots of the kinematic invariants:
\begin{equation}
\begin{split}
q_1&=\sqrt{y (y-4)}, \hspace{0.5cm} q_2=\sqrt{z (z-4)}, \hspace{0.5cm} q_3=\sqrt{y (y+4)}, \hspace{0.5cm} q_4=\sqrt{y z (y (z-4)-4 z)},\\
q_5&=\sqrt{y (y+z) (-4 z+y (y+z))}, \hspace{0.5cm} q_6=\sqrt{z (y+z) \left(y (z-4)+z^2\right)}, \hspace{0.5cm} q_7=\sqrt{(y+z) (y+z+4)},\\
q_8&=\sqrt{-y z (y+z)}, \hspace{0.5cm} q_9=\sqrt{4-z}.
\end{split}
\end{equation}
The subset (II), which is associated to MIs whose analytic structure involve elliptic functions, contains two sectors (see fig. \ref{fig:elliptic_sec}). The first one is a non-planar triangle with a massive loop \cite{vonManteuffel:2017hms}, shown in fig.~\ref{fig:elliptic_sec}~(a).
This sector has 2 MIs, which admit a representation in terms of \emph{elliptic multiple polylogarithms} (eMPLs) \cite{Broedel:2019hyg}.

We choose for them the same normalization as in \cite{vonManteuffel:2017hms}:
\begin{equation}
g_{31}(y)=y^2 \epsilon ^4 
\mathcal{J}_{19} \, ,
\hspace{0.5cm} g_{32}(y)=-\frac{y^2 \epsilon ^4 \left(16 +y\right) }{2(1 + 2\epsilon)}
\mathcal{J}_{20} \, .
\end{equation}
The second sector whose analytic structure is characterised by the presence of elliptic functions is the top sectors of the topology NPL, i.e. the double-box integral, shown in fig.~\ref{fig:elliptic_sec}~(b), which contains 4 MIs: $g_{33},g_{34},g_{35}$ and $g_{36}$. We choose, as basis for this sector, the following scalar integrals:
\begin{equation}
\begin{split}
g_{33}(y,z)&=-y^3 \epsilon ^4 \mathcal{J}_{39}, \hspace{0.5cm} 
g_{34}(y,z)=y^2 \epsilon ^4 \mathcal{J}_{40} \, , \\ 
g_{35}(y,z)&=y^4 \epsilon ^4 \mathcal{J}_{41} \, , \hspace{0.5cm} 
g_{36}(y,z)=-y \epsilon ^4 \mathcal{J}_{42} \, .
\end{split}
\end{equation}
With this choice of normalization, initial conditions in the origin $y=z=0$ are:
\begin{equation}
    g_1 = f_1, \hspace{0.5cm} g_2(y) = f_2(y), \hspace{0.5cm} g_i(y=0,z=0)=0, \hspace{0.5cm} i=3,\dots,36
\end{equation}
where $f_1$ and $f_2$ are given in Eq.~(\ref{f1f2}).

The non-polylogarithmic structure of this sector is two-fold. First, the differential equations for the MIs $g_{33}$, $g_{34}$, $g_{35}$ and $g_{36}$ contain the triangle integrals $g_{31}$ and $g_{32}$ in the non-homogeneous part of the system. As a consequence the analytic solution of the differential equations requires the integration over kernels that contains eMPLs. Moreover, also the homogeneous part of the differential equations itself contains elliptic functions. This statement can be verified by studying the maximal cut of the double-box integral \cite{Henn:2014qga}:
\begin{equation}
\mathcal{I}_{db} = \mathcal{J}_{39} \, .
\label{eq:db}
\end{equation}
In order to perform such computation it is convenient to follow the loop-by-loop analysis \cite{Frellesvig:2017aai,Harley:2017qut} of the Baikov \cite{Baikov:1996cd} representation of the integral $\mathcal{I}_{db}$.
For reader's convenience, we briefly introduce the basic idea of Baikov representation and its application to maximal cut analysis, and we refer to refs.~\cite{Baikov:1996cd,Frellesvig:2017aai,Harley:2017qut,Dlapa:2022nct} for a proper description of the method. The Baikov representation of an $L$-loop Feynman integral in $D$ dimensions, with E independent external momenta, is written as:
\begin{equation} \label{eq:baikov}
    \mathcal{I}(a_1,\cdots,a_n) = C_E^L \int \frac{dz_1 \cdots dz_n}{z_1^{a_1} \cdots z_n^{a_n}} U^{\frac{E-D+1}{2}} P^{\frac{D-L-E-1}{2}}.
\end{equation}
The variables $z_i$ are just the scalar products $D_i$ which define the integral topology. Indeed, the basic idea of the Baikov representation is to write a scalar Feynman integral in a form where the integration variables are exactly the scalar products associated to the topology. 
The factors U and P are, respectively, the Gram determinant of the external momenta $p_1,\cdots,p_E$:
\begin{equation}
    U = \det G(p_1,\cdots,p_E),
\end{equation}
and the Gram determinant of the external momenta together with the loop ones:
\begin{equation}
    P = \det G(k_1,\cdots,k_L, p_1, \cdots, p_E),
\end{equation}
where $G_{ij}(\vec{v}) = v_i \cdot v_j$ is the Gram matrix. The factor $C_E^L$ is a normalization constant, which is immaterial for the present discussion.

In the Baikov representation the maximal cut of the integral $\mathcal{I}(a_1,\cdots,a_n)$ can be calculated from the multivariate residues of the expression \eqref{eq:baikov} \cite{Cachazo:2008vp}. This operation is more efficiently done within the so-called \emph{loop-by-loop approach} \cite{Frellesvig:2017aai}. In a nutshell, in this approach we split the integral under study in subsequent one-loop integrals and then we take the maximal cut of each sub-integral sequentially. Let us consider explicitly our double box integral \eqref{eq:db}. By inspecting the integrand of \eqref{eq:db} we see that it can be split into two one-loop box integrals as follows:
\begin{equation} 
\label{eq:baikov_db}
    \mathcal{I}_{db} = \mathcal{C} \int \frac{dz_1 dz_2 dz_3 dz_8}{z_1 z_2 z_3 z_8} U_1^{\epsilon} P_1 ^{-\frac{1}{2}-\epsilon} \int \frac{dz_4 dz_5 dz_6 dz_7}{z_4 z_5 z_6 z_7} U_2^{\epsilon} P_2 ^{-\frac{1}{2}-\epsilon},
\end{equation}
where $\mathcal{C}$ is some overall normalisation constant. $U_2$ and $P_2$ are Gram determinants associated to the change of variables for a one-loop box integral with loop momentum $k_2$, similarly $U_1$ and $P_1$ are related to change of variables for a box integral with loop momentum $k_1$. Then, the maximal cut of the expression \eqref{eq:baikov_db} is obtained by taking the residue around the simple pole $z_1 = \cdots z_7 = 0$:
\begin{align} \label{eq:db_maxcut}
    \operatorname{MCut} \, \mathcal{I}_{db} & =\mathcal{C} \oint_{z_1=z_2=z_3=0} \frac{dz_1 dz_2 dz_3 dz_8}{z_1 z_2 z_3 z_8} U_1^{\epsilon} P_1 ^{-\frac{1}{2}-\epsilon} \oint_{z_4=z_5=z_6=z_7=0} \frac{dz_4 dz_5 dz_6 dz_7}{z_4 z_5 z_6 z_7} U_2^{\epsilon} P_2 ^{-\frac{1}{2}-\epsilon} \nn \\
    & = \mathcal{C} \int dz_8 \frac{1}{s\,z_8\sqrt{(z_8 + t)(z_8 + s + t)(z_8 - z_{+})(z_8 - z_{-})}}, 
\end{align}
where $z_{\pm} = \frac{1}{2}\left(-s - 2t \pm \sqrt{s}\sqrt{s + 16m_t^2}\right)$. As we can see from Eq.~\eqref{eq:db_maxcut} the result of the maximal cut for the double box non planar integral in fig.~\ref{fig:elliptic_sec}~(b) is a one-fold integral. It is possible to show that Eq.~\eqref{eq:db_maxcut} can be written in terms of complete elliptic integrals of first and third kind, where the elliptic curve is given by the polynomial of fourth-order in the integration variable $z_8$:
\begin{equation} \label{eq:ell_curve}
    y^2_c = (z_8 + t)(z_8 + s + t)(z_8 - z_{+})(z_8 - z_{-}).
\end{equation}
Remarkably, we observe that the elliptic curve \eqref{eq:ell_curve} degenerates to the same curve of the massive triangle in fig.~\ref{fig:elliptic_sec}~(a) \cite{vonManteuffel:2017hms,Broedel:2019hyg,Abreu:2022vei} in the forward limit $t = 0$. 

\subsection{Semi-Analytic solution with Generalised power series}

We conclude this section by describing the solution for the systems of differential equations associated to the planar topology PLA and the non-planar topology NPL. As already mentioned, we choose to exploit the generalised power series method, as described in \cite{Moriello:2019yhu} and implemented in the software \texttt{DiffExp} \cite{Hidding:2020ytt}, to obtain a semi-analytic solution for the set of MIs. This method has the advantage of not being limited by the functional space in which the MIs would be analytically represented. This feature allows us to avoid the issues connected with the presence of MIs which admit an analytic solution in terms of elliptic integrals, for which both the understanding of their analytic structure and the numerical evaluation can still represent a bottleneck for phenomenological applications.

We exploit the method to build a grid of points for the contribution of the corrections considered in this paper to the hard function $\mathcal{H}^{\gamma\gamma}_{NNLO}$. After interpolation, the grid has been used in the fully massive NNLO phenomenological study for diphoton production in \cite{Becchetti:ta}. 
The grid has been generated directly in the physical region of the phase-space for this process:
\begin{equation}
    s > 0, \;\;\; t=-\frac{s}{2}(1-\cos(\theta)), \;\;\; -s < t <0,
\end{equation}
\begin{figure}[h!]
\centering
\includegraphics[width =7 cm]{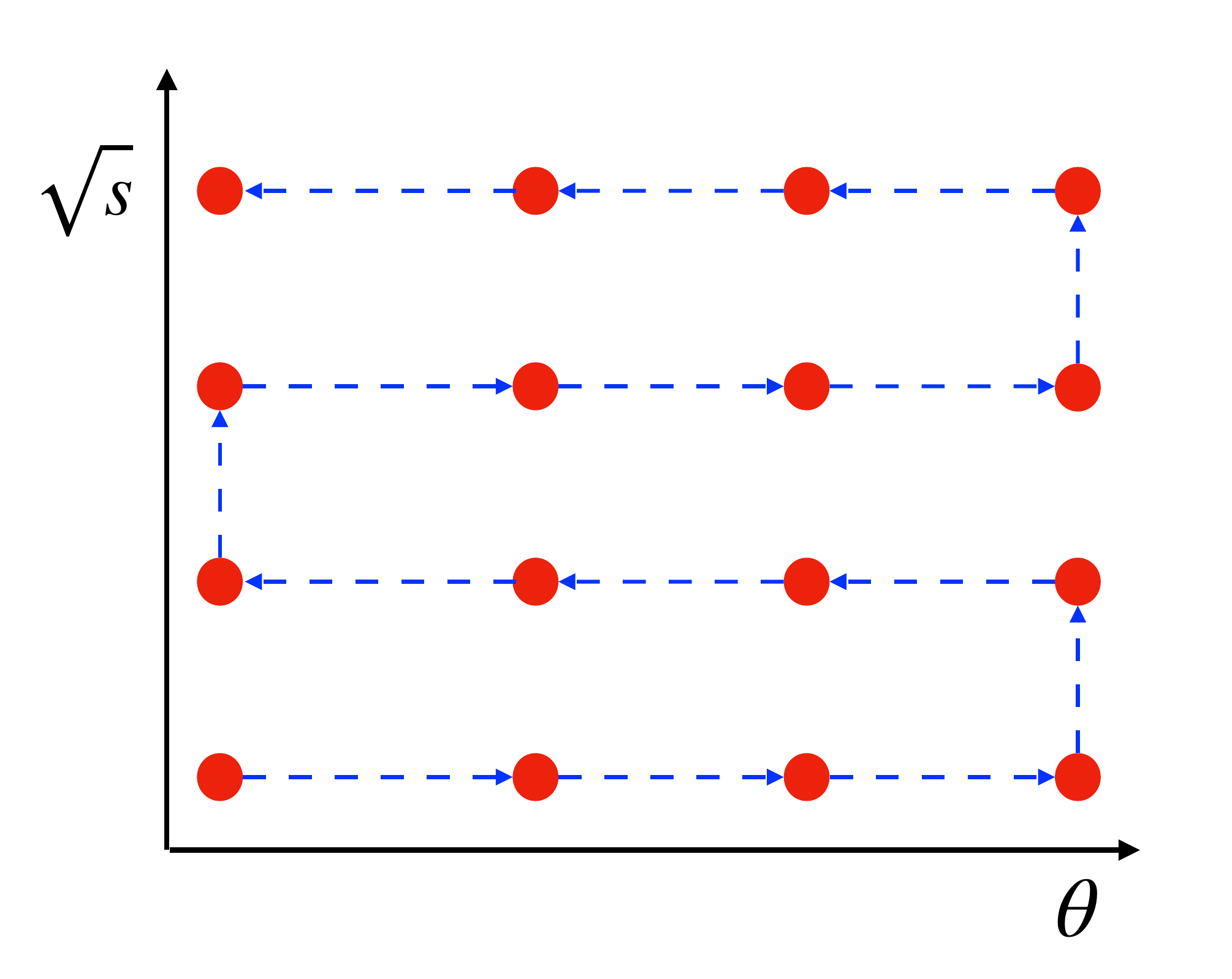}
\caption{Schematic representation of the procedure exploited to optimise the grid construction within \texttt{DiffExp}. Red dots represents the points in which the MIs are evaluated and the blue dashed lines connect the sequential evaluations.} 
\label{fig:grid}
\end{figure}
where $\theta$, $0 < \theta < \pi$, is the scattering angle in the partonic center of mass frame. 

Since the evaluation time, within \texttt{DiffExp}, for the MIs needed in this process is relatively low, we can build the grid of points as follows\footnote{We notice that for more CPU demanding computations more refined approaches have to be used.}.
We consider a total number of 13752 points in the following range for the scattering angle $\theta$ and the energy of the center of mass $s$:
\begin{equation}
    -0.99 < \cos{\theta} < 0.99, \;\;\; 8 \operatorname{GeV} < \sqrt{s} < 2.2 \operatorname{TeV}.
\end{equation}

The points $p_{i,j}$ of the grid are equally spaced in the coordinate system $(s,t)$\footnote{For the purpose of this discussion we use the dimensional Mandelstam variables defined in Eq. \ref{eq:mandelstam}} as follows:
\begin{equation}
    p_{i,j} := \begin{cases} s_i = s_0 + (s_f - s_0)\frac{i}{572} 
    \\ t_j = -\frac{s_i}{2}(1- \cos{\theta_j}), \;\;\; \cos{\theta_j} = \cos{\theta_0} + (\cos{\theta_f}-\cos{\theta_0})\frac{j}{23}  
    \end{cases}
\end{equation}
where $s_0 = 64 \operatorname{GeV}^2$, $s_f = 4.4 \operatorname{TeV}^2$, $\cos{\theta_0} = -0.99$, $\cos{\theta_f} = 0.99$ and the indices $i,j$ take the values $i\in [0,572]$, $j \in [0,23]$. We constructed the grid by performing sequential numerical evaluations of the MIs in \texttt{DiffExp} as depicted schematically in figure \ref{fig:grid}.
Starting from the boundary conditions for the systems of differential equations, we perform a first evaluation in the physical point $p_{0,0}$. From this point, at fixed value of $s$, we move along the $\theta$ axis, from $\cos{\theta} = -0.99$ to $\cos{\theta} = 0.99$, up to the point $p_{0,23}$. Then, we increase the value of $s$ and we move in the other direction along the $\theta$ axis up to the point $p_{1,0}$, and so on so forth. In order to optimise the grid generation, for each evaluation we use as boundary conditions the value of the MIs obtained at the previous point. This procedure effectively increases the efficiency of the evaluation for the MIs. In particular, we managed to evaluate the MIs in all the points of the grid, with a 16 digits accuracy, in 2.5 hours for the system PLA and 10.5 hours for the system NPL, on a single core laptop.

\begin{figure}[h!] 
\centering

     \includegraphics[width=6cm]{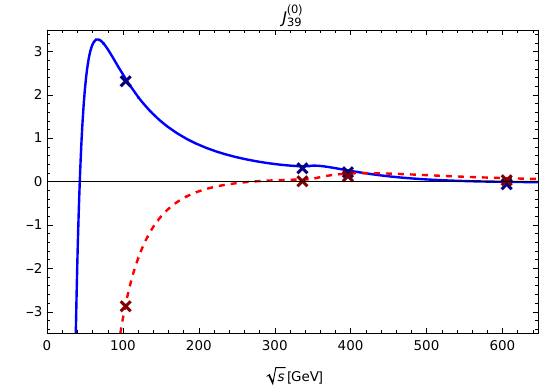}
\hfill
     \includegraphics[width=6cm]{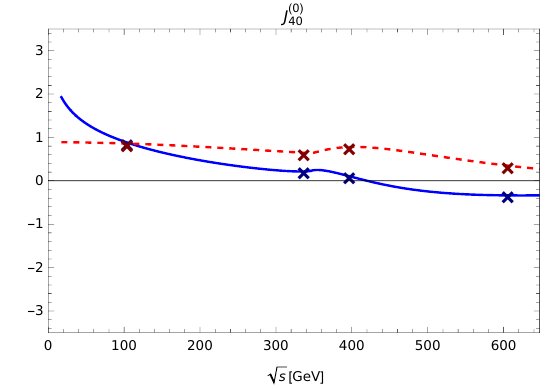}
 
 \medskip
     \includegraphics[width=6cm]{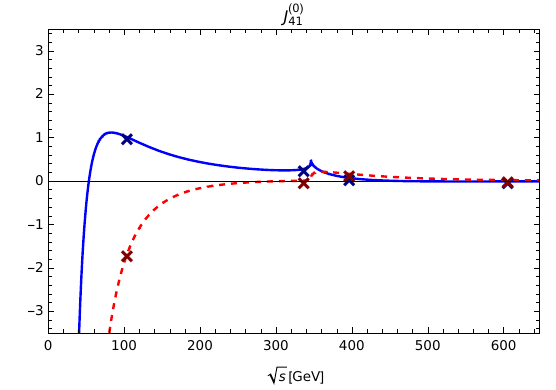}
 \hfill
     \includegraphics[width=6cm]{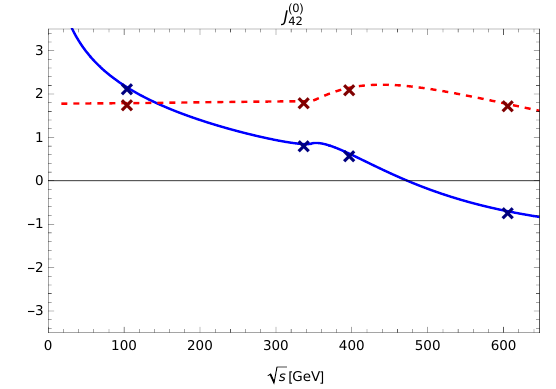}

\caption{Numerical checks for the non-planar double-box MIs $\mathcal{J}_{39},\mathcal{J}_{40},\mathcal{J}_{41}$ and $\mathcal{J}_{42}$. The plots show the order $\mathcal{O}(\epsilon^0)$ of the masters. Blue and dashed red lines represent, respectively, the numerical values of real and imaginary part of the MIs obtained with \texttt{DiffExp}. Crossed dots represent numerical values obtained with \texttt{AMFlow}. The evaluations are performed for different values of $\sqrt{s}$ at fixed angle $\theta$ (in this case $\cos{\theta}=0.7$).} 
\label{fig:checks}

\end{figure}

Finally, in order to validate our results, we performed numerical checks for the MIs against independent numerical evaluations done with the \texttt{AMFlow} package \cite{Liu_2023}, which implements the auxiliary mass flow method \cite{Liu:2017jxz,Liu:2021wks}. The MIs have been checked for several points in the physical phase-space region, finding an agreement between the two independent evaluations up to 200 digits of accuracy. As a proof of concept of the numerical checks we show in figure \ref{fig:checks} our results for the double-box MIs in the non-planar topology NPL.

\section{Conclusions}

In this paper we presented the computation of the two-loop form factors for diphoton production in the $q\bar{q}$ channel, where the full dependence on the top quark mass has been retained. This computation represents the only missing ingredient, at two loops, in order to be able to perform a phenomenological study for diphoton production at NNLO \cite{Becchetti:ta} which fully takes into account the dependence on the top quark mass in all the relevant channels. 

The non-planar topology which contributes to this process contains two sectors of MIs whose analytic representation cannot be given in terms of MPLs. In order to be able to exploit our results for phenomenological applications, we computed the MIs by means of differential equations, exploiting the generalised power series technique. This method proves to be of great use for phenomenological applications, especially in cases where the functional space for the MIs contains not only polylogarithmic functions.

\section*{Acknowledgements}
This work is supported by the Spanish Government (Agencia Estatal de Investigaci\'on MCIN/AEI/ 10.13039/501100011033) Grant No. PID2020-114473GB-I00, and Generalitat Valenciana Grants No. PROMETEO/2021/071 and ASFAE/2022/009 (Planes Complementarios de I+D+i, Next Generation EU). 
M.B. acknowledges the financial support from the European Union Horizon 2020 research and innovation programme: High precision multi-jet dynamics at the LHC (grant agreement no. 772009). L.C. and F.C. are supported by Generalitat Valenciana GenT Excellence Programme (CIDEGENT/2020/011) and ILINK22045.

\appendix

\section{The master integrals}
\label{appendix}
The MIs appearing in the form factors, modulo permutations of the external momenta, are the following:
\begin{align*}
\mathcal{J}_1 &= \mathcal{I}_{\text{PLA}}(0, 0, 0, 2, 0, 0, 2, 0, 0), \hspace{1cm}
\mathcal{J}_2(y) = \mathcal{I}_{\text{PLA}}(0, 1, 2, 0, 0, 0, 2, 0, 0), \\
\mathcal{J}_3(y) &= \mathcal{I}_{\text{PLA}}(0, 0, 0, 2, 1, 2, 0, 0, 0), \hspace{1cm}
\mathcal{J}_4(y) = \mathcal{I}_{\text{PLA}}(0, 0, 2, 1, 2, 0, 0, 0, 0), \\
\mathcal{J}_5(y) &= \mathcal{I}_{\text{PLA}}(0, 0, 1, 2, 2, 0, 0, 0, 0), \hspace{1cm} 
\mathcal{J}_6(y) = \mathcal{I}_{\text{PLA}}(0, 1, 2, 0, 2, 1, 0, 0, 0), \\
\mathcal{J}_7(y) &= \mathcal{I}_{\text{PLA}}(0, 0, 0, 2, 1, 1, 1, 0, 0), \hspace{1cm} 
\mathcal{J}_8(y) = \mathcal{I}_{\text{PLA}}(0, 0, 1, 3, 1, 0, 1, 0, 0), \\
\mathcal{J}_9(y) &= \mathcal{I}_{\text{PLA}}(0, 1, 1, 2, 0, 0, 1, 0, 0), \hspace{1cm} 
\mathcal{J}_{10}(y) = \mathcal{I}_{\text{PLA}}(0, 1, 2, 1, 0, 0, 2, 0, 0), \\
\mathcal{J}_{11}(y) &= \mathcal{I}_{\text{PLA}}(1, 0, 0, 2, 1, 1, 0, 0, 0), \hspace{1cm} 
\mathcal{J}_{12}(y) = \mathcal{I}_{\text{PLA}}(2, 0, 0, 1, 1, 1, 0, 0, 0), \\
\mathcal{J}_{13}(y) &= \mathcal{I}_{\text{PLA}}(1, 0, 0, 2, 1, 2, 0, 0, 0), \hspace{1cm} 
\mathcal{J}_{14}(y) = \mathcal{I}_{\text{PLA}}(0, 1, 1, 1, 1, 1, 0, 0, 0), \\
\mathcal{J}_{15}(y) &= \mathcal{I}_{\text{PLA}}(0, 2, 1, 0, 1, 1, 1, 0, 0), \hspace{1cm} 
\mathcal{J}_{16}(y) = \mathcal{I}_{\text{PLA}}(0, 1, 1, 1, 0, 1, 1, 0, 0), \\
\mathcal{J}_{17}(y) &= \mathcal{I}_{\text{NPL}}(0, 1, 0, 1, 1, 1, 1, 0, 0), \hspace{1cm} 
\mathcal{J}_{18}(y) = \mathcal{I}_{\text{PLA}}(0, 1, 1, 1, 1, 1, 1, 0, 0), \\
\mathcal{J}_{19}(y) &= \mathcal{I}_{\text{NPL}}(0, 1, 1, 1, 1, 1, 1, 0, 0), \hspace{1cm} 
\mathcal{J}_{20}(y) = \mathcal{I}_{\text{NPL}}(0, 1, 1, 2, 1, 1, 1, 0, 0), \\
\mathcal{J}_{21}(y,z) &= \mathcal{I}_{\text{PLB}}(1, 1, 1, 2, 0, 1, 0, 0, 0), \hspace{1cm} 
\mathcal{J}_{22}(y,z) = \mathcal{I}_{\text{PLA}}(1, 1, 1, 2, 0, 0, 1, 0, 0), \\
\mathcal{J}_{23}(y,z) &= \mathcal{I}_{\text{PLA}}(1, 1, 1, 2, 0, 0, 1, -1, 0), \hspace{0.7cm} 
\mathcal{J}_{24}(y,z) = \mathcal{I}_{\text{PLA}}(1, 0, 1, 1, 1, 0, 1, 0, 0), \\
\mathcal{J}_{25}(y,z) &= \mathcal{I}_{\text{PLA}}(1, 0, 1, 2, 1, 0, 1, 0, 0), \hspace{1cm} 
\mathcal{J}_{26}(y,z) = \mathcal{I}_{\text{PLA}}(1, 0, 0, 2, 1, 1, 1, 0, 0), \\
\mathcal{J}_{27}(y,z) &= \mathcal{I}_{\text{PLA}}(1, 0, 0, 3, 1, 1, 1, 0, 0), \hspace{1cm} 
\mathcal{J}_{28}(y,z) = \mathcal{I}_{\text{PLA}}(1, 0, 0, 2, 1, 1, 1, 0, -1), \\
\mathcal{J}_{29}(y,z) &= \mathcal{I}_{\text{PLA}}(1, 1, 1, 1, 1, 0, 1, 0, 0), \hspace{1cm} 
\mathcal{J}_{30}(y,z) = \mathcal{I}_{\text{NPL}}(1, 0, 0, 1, 1, 1, 1, 0, 0), \\
\mathcal{J}_{31}(y,z) &= \mathcal{I}_{\text{NPL}}(2, 0, 0, 1, 1, 1, 1, 0, 0), \hspace{1cm} 
\mathcal{J}_{32}(y,z) = \mathcal{I}_{\text{NPL}}(1, 0, 0, 1, 2, 1, 1, 0, 0), \\
\mathcal{J}_{33}(y,z) &= \mathcal{I}_{\text{NPL}}(1, 1, 0, 1, 1, 1, 1, 0, 0), \hspace{1cm} 
\mathcal{J}_{34}(y,z) = \mathcal{I}_{\text{NPL}}(1, 1, 0, 2, 1, 1, 1, 0, 0), \\
\mathcal{J}_{35}(y,z) &= \mathcal{I}_{\text{PLA}}(1, 1, 1, 1, 1, 1, 1, 0, 0), \hspace{1cm} 
\mathcal{J}_{36}(y,z) = \mathcal{I}_{\text{PLA}}(1, 1, 1, 1, 1, 1, 1, -1, 0), \\
\mathcal{J}_{37}(y,z) &= \mathcal{I}_{\text{PLA}}(1, 1, 1, 1, 1, 1, 1, 0, -1), \hspace{0.7cm} 
\mathcal{J}_{38}(y,z) = \mathcal{I}_{\text{PLA}}(1, 1, 1, 1, 1, 1, 1, -1, -1), \\
\mathcal{J}_{39}(y,z) &= \mathcal{I}_{\text{NPL}}(1, 1, 1, 1, 1, 1, 1, 0, 0), \hspace{1cm} 
\mathcal{J}_{40}(y,z) = \mathcal{I}_{\text{NPL}}(1, 1, 1, 1, 1, 1, 1, -1, 0), \\
\mathcal{J}_{41}(y,z) &= \mathcal{I}_{\text{NPL}}(1, 1, 1, 1, 2, 1, 1, 0, 0), \hspace{1cm} 
\mathcal{J}_{42}(y,z) = \mathcal{I}_{\text{NPL}}(1, 1, 1, 1, 1, 1, 1, -1, -1).
\end{align*}
To complete the whole set of MIs appearing in the form factors, we have to consider also the following masters, obtained from the previous list permuting the kinematic variables ($y$,$z$ and $w=-y-z$):
\begin{align*}
\mathcal{J}_{43}(z) & = \mathcal{J}_{2}(z), \hspace{1.6cm}
\mathcal{J}_{44}(w)  = \mathcal{J}_{2}(w), \\
\mathcal{J}_{45}(w) & = \mathcal{J}_{4}(w), \hspace{1.6cm}
\mathcal{J}_{46}(z)  = \mathcal{J}_{4}(z), \\
\mathcal{J}_{47}(w) & = \mathcal{J}_{5}(w), \hspace{1.6cm}
\mathcal{J}_{48}(z)  = \mathcal{J}_{5}(z), \\
\mathcal{J}_{49}(w) & = \mathcal{J}_{8}(w), \hspace{1.6cm}
\mathcal{J}_{50}(z)  = \mathcal{J}_{8}(z), \\
\mathcal{J}_{51}(y,z) & = \mathcal{J}_{21}(y,w), \hspace{1cm}
\mathcal{J}_{52}(y,z)  = \mathcal{J}_{22}(y,w), \\
\mathcal{J}_{53}(y,z) & = \mathcal{J}_{23}(y,w), \hspace{1cm}
\mathcal{J}_{54}(y,z)  = \mathcal{J}_{24}(y,w), \\
\mathcal{J}_{55}(y,z) & = \mathcal{J}_{25}(y,w), \hspace{1cm}
\mathcal{J}_{56}(y,z)  = \mathcal{J}_{26}(y,w), \\
\mathcal{J}_{57}(y,z) & = \mathcal{J}_{27}(y,w), \hspace{1cm}
\mathcal{J}_{58}(y,z)  = \mathcal{J}_{28}(y,w), \\
\mathcal{J}_{59}(y,z) & = \mathcal{J}_{29}(y,w), \hspace{1cm}
\mathcal{J}_{60}(y,z)  = \mathcal{J}_{30}(y,w), \\
\mathcal{J}_{61}(y,z) & = \mathcal{J}_{31}(y,w), \hspace{1cm}
\mathcal{J}_{62}(y,z)  = \mathcal{J}_{32}(y,w), \\
\mathcal{J}_{63}(y,z) & = \mathcal{J}_{33}(y,w), \hspace{1cm}
\mathcal{J}_{64}(y,z)  = \mathcal{J}_{34}(y,w), \\
\mathcal{J}_{65}(y,z) & = \mathcal{J}_{35}(y,w), \hspace{1cm}
\mathcal{J}_{66}(y,z)  = \mathcal{J}_{36}(y,w), \\
\mathcal{J}_{67}(y,z) & = \mathcal{J}_{37}(y,w), \hspace{1cm}
\mathcal{J}_{68}(y,z)  = \mathcal{J}_{38}(y,w), \\
\mathcal{J}_{69}(y,z) & = \mathcal{J}_{39}(y,w), \hspace{1cm}
\mathcal{J}_{70}(y,z)  = \mathcal{J}_{40}(y,w), \\
\mathcal{J}_{71}(y,z) & = \mathcal{J}_{41}(y,w), \hspace{1cm}
\mathcal{J}_{72}(y,z)  = \mathcal{J}_{42}(y,w).
\end{align*}

\newpage

\bibliographystyle{JHEP}
\bibliography{biblio}

\end{document}